\documentclass[prl,twocolumn,amsfonts,nofootinbib,showpacs]{revtex4}
\usepackage{epsfig,bm,amsmath}
\usepackage[latin1]{inputenc}
\usepackage[T1]{fontenc}

\newcommand{\xbf}{\bm{x}}
\newcommand{\pbf}{\bm{p}}

\begin{document}
\preprint{}
\title{Interpretation of neutrino oscillations based on new physics in the infrared}
\author{J.M. Carmona}
\email{jcarmona, cortes, indurain@unizar.es}
\affiliation{Departamento de F\'{\i}sica Te\'orica,
Universidad de Zaragoza, Zaragoza 50009, Spain}
\author{J.L. Cort\'es}
\email{jcarmona, cortes, indurain@unizar.es}
\affiliation{Departamento de F\'{\i}sica Te\'orica,
Universidad de Zaragoza, Zaragoza 50009, Spain}
\author{J. Indur\'ain}
\email{jcarmona, cortes, indurain@unizar.es}
\affiliation{Departamento de F\'{\i}sica Te\'orica,
Universidad de Zaragoza, Zaragoza 50009, Spain}
\begin{abstract}
An interpretation of neutrino oscillations based on a modification of
relativistic quantum field theory at low energies, without the need to
introduce a neutrino mass, is seen to be compatible with all observations.
\end{abstract}
\pacs{14.60.Pq, 11.30.Cp, 11.10.-z} 
\maketitle

\section{Introduction}

The now well-established observation of deficit of solar neutrinos,
atmospheric neutrinos, and neutrinos from reactors and accelerators finds
a coherent interpretation in terms of neutrino
oscillations between three neutrino flavors of different masses~\cite{oscillations}.
In the minimal standard model (SM), and in contrast with the rest
of the matter particles, the neutrino is assumed to be a zero mass,
left-handed fermion. Therefore neutrino oscillations is our first
glimpse of physics beyond the SM.

Massive neutrinos are introduced in extensions of the SM which
normally invoke new physics at high energies. In particular, one can
consider a Majorana mass term for the neutrino, generated by a
five-dimensional operator in the SM Lagrangian which would be
suppressed by the inverse of a certain high-energy scale.
Another possibility is to enlarge the field content of the SM with
a right-handed neutrino, which allows mass to be generated by
the usual Higgs mechanism. One has to account however for the smallness
of the neutrino mass, which is achieved by the see-saw mechanism~\cite{see-saw},
again invoking a grand-unification scale.

The presence of new physics at high energies has been explored in
several attempts to find alternatives to the standard neutrino
oscillation mechanism.
This new physics might include Lorentz and/or CPT
violations. These two low-energy symmetries are being questioned at
very high energies in the framework of quantum gravity and string theory
developments~\cite{qugr}, and in fact simple models with Lorentz and/or
CPT violations are able to generate neutrino oscillations,
even for massless neutrinos~\cite{coleman}. Some of them are considered in
the context of the Standard Model Extension (SME)~\cite{kostel}, which
is the most general framework for studying Lorentz and CPT violations
in effective field theories.

However, all these alternative mechanisms involve new energy dependencies
of the oscillations which are in general disfavored over the standard
oscillation mechanism by experimental data,
which also put strong bounds on the contribution of new physics to this
phenomenon~\cite{fits}. This seems to indicate that our understanding
of neutrino oscillations as driven by mass differences between neutrino flavors
is indeed correct.

In this letter we want to argue that this might not be the case. We will
present an example of new physics to the SM able to generate neutrino
oscillations and which is essentially different from previously considered
models in one or several of the following aspects: it does not necessarily
add new fields to those present in the SM, it may be completely indistinguishable
from the standard oscillation mechanism in the energy ranges where the
phenomenon has been studied (for neutrinos of medium and high energies), so
that automatically satisfies all constraints which are already fulfilled by
the standard mechanism, and finally, it predicts new physics in the infrared,
so that future neutrino low energy experiments could
distinguish this mechanism from the standard one.

Our example will be based on the so-called theory of noncommutative
quantum fields, which has recently been proposed
as an specific scheme going beyond quantum field
theory~\cite{ncqft1,ncqft2,ncqftother}.
The consequences for neutrino oscillations of a simple model with
modified anticommutators for the neutrino fields, which can be
identified as an example of a SME in the neutrino sector, has very
recently been explored in Ref.~\cite{gamboa}.
We will see however that it is possible to introduce a generalization
of the anticommutation relations of fields in a more general way than
that studied in Ref.~\cite{gamboa}, going beyond the effective field
theory framework of the SME,
which is the key to reproduce the oscillation results without the
need to introduce a neutrino mass, and with new consequences at
infrared energies.

\section{Noncanonical fields and neutrino oscillations}

The theory of noncommutative fields was first considered in
Refs.~\cite{ncqft1,ncqft2}. It is an extension of the usual canonical
quantum field theory in which the procedure of quantification of a
classical field theory is changed in the following way:
the quantum Hamiltonian remains the same
as the classical Hamiltonian, but the canonical commutation
relations between fields are modified. In the case of the scalar
complex field this modification leads to the introduction of two
new energy scales (one infrared or low-energy scale, and another one
ultraviolet or high-energy scale), together with new observable
effects resulting from the modification of the dispersion relation
of the elementary excitations of the fields~\cite{ncqft2}. If one is
far away from any of these two scales the theory approaches the
canonical relativistic quantum field theory with corrections involving
Lorentz invariance violations which can be expanded in powers of the
ratios of the infrared scale over the energy and the energy over the
ultraviolet scale. By an appropriate choice of the two new energy
scales one can make the departures from the relativistic theory
arbitrarily small in a certain energy domain.

We will now explore the relevance of the extension of relativistic
quantum field theory based on noncanonical fields in neutrino
oscillations. In particular, we will show that it is possible to
obtain oscillations with the observed experimental properties just
by considering a modification of the anticommutators of the fields
appearing in the SM, without the need to introduce a right-handed
neutrino or a mass for this particle.

With the left-handed lepton fields of the SM
\begin{equation}
\Psi_{L \alpha}=\left(\begin{array}{c}\nu_{\alpha} \\
  l_{\alpha}\end{array}\right)_L ,
\end{equation}
where $\alpha$ runs the flavor indices, $(\alpha=e,\mu,\tau)$, the
simplest way to consider an analog of the extension of the canonical
quantum field theory for a complex scalar field proposed in
Refs.~\cite{ncqft1,ncqft2} is to introduce the modified anticommutation
relations
\begin{equation}
\begin{split}
\{\nu_{L\alpha}(\bm{x}),\nu^{\dagger}_{L\beta}(\bm{y})\}&=
\{l_{L\alpha}(\bm{x}),l^{\dagger}_{L\beta}(\bm{y})\}= \\
&\left[\delta_{\alpha\beta} + A_{\alpha\beta}\right]
\,\delta^3(\bm{x}-\bm{y}).
\label{leptoncomm}
\end{split}
\end{equation}
A particular choice for the matrix $A_{\alpha\beta}$ in flavor space which
parametrizes the departure from the canonical anticommutators
corresponds to the new mechanism for neutrino oscillations proposed
in Ref.~\cite{gamboa} which, however, is not compatible with the
energy dependence of the experimental data.

In order to reproduce the observed properties of neutrino
oscillations~\cite{oscillations} one has to go beyond this extension
and consider an anticommutator between fields at different
points. This can be made compatible with rotational and translational
invariance and with $SU(2)\times U(1)_Y$ gauge symmetry by making use
of the Higgs field
\begin{equation}
  \Phi=\left(\begin{array}{c}\varphi^+ \\
  \varphi^0 \end{array}\right),
  \tilde{\Phi}=\left(\begin{array}{c}\varphi^{0*} \\
  -\varphi^- \end{array}\right).
\end{equation}
The modified anticommutators of the left-handed lepton fields that we
consider in this work are
\begin{equation}
\begin{split}
\{\Psi_{L\alpha}(\bm{x}), (\Psi_{L\beta})^\dagger(\bm{y})\}&=
\delta_{\alpha\beta}\,\delta^3(\bm{x}-\bm{y}) \\
&+ \tilde{\Phi}(\bm{x})
\tilde{\Phi}^\dagger(\bm{y}) B_{\alpha\beta}(|\bm{x}-\bm{y}|),
\label{phicomm}
\end{split}
\end{equation}
where $B_{\alpha\beta}$ are now functions of $|\bm{x}-\bm{y}|$ instead
of constants. Note that Eq.~(\ref{phicomm}) is compatible with gauge 
invariance since $\tilde{\Phi}(\bm{x})$ has the same $SU(2)\times U(1)_Y$
quantum numbers as $\Psi_{L\alpha}(\bm{x})$.

After introduction of spontaneous symmetry breaking ($\langle
\varphi^0 \rangle = v/\sqrt{2}$),
and neglecting effects coming from the fluctuation of the scalar field
which surely is a good approximation for neutrino oscillations, the
only anticommutators that are changed are those of the neutrino fields
\begin{equation}
\{\nu_{L\alpha}(\bm{x}),\nu^{\dagger}_{L\beta}(\bm{y})\}=
\delta^3(\bm{x}-\bm{y})\,\delta_{\alpha\beta}+C_{\alpha\beta}(|\bm{x}-\bm{y}|),
\label{neutrinocomm}
\end{equation}
where
\begin{equation}
C_{\alpha\beta}(|\bm{x}-\bm{y}|) = \frac{v^2}{2}
B_{\alpha\beta}(|\bm{x}-\bm{y}|).
\end{equation}

One can suspect that the new anticommutation relations Eq.~(\ref{neutrinocomm})
introduce a source of mixing between flavors that will affect neutrino oscillations.
We will see in the next Section that this is indeed the case.

\section{Solution of the free theory}

In order to study the neutrino oscillations induced by the modified
anticommutators of fields in the neutrino sector, 
one needs to solve the free theory given by the Hamiltonian
\begin{equation}
H=\sum_{\alpha}\int d^3\bm{x} \left[i\,\nu_{L\alpha}^\dagger
\left(\bm{\sigma}\cdot\bm{\nabla}\right) \nu_{L\alpha}\right]
\label{ham}
\end{equation}
(where $\bm{\sigma}$ are the $2\times 2$ Pauli matrices),
and the anticommutation relations
showed in Eq.~(\ref{neutrinocomm}).

Let us introduce a plane wave expansion for the neutrino field
\begin{equation}
\begin{split}
\nu_{L\alpha} (\xbf) =& \int \frac{d^3\pbf}{(2\pi)^3} \, \frac{1}{\sqrt{2p}}\,
\sum_{i} \left[\, b_i(\pbf)\, u_{L\alpha}^i (\pbf) \, e^{i
    \pbf\cdot\xbf}\right. \\ +&
\,\left. d_i^\dagger(\pbf)\, v_{L\alpha}^i (\pbf) \, e^{-i \pbf\cdot\xbf} \right],
\label{neutrino}
\end{split}
\end{equation}
where $p=|\pbf|$, and $(b_i(\pbf),d_i^\dagger(\pbf))$ are the annihilation
and creation operators  of three types of particles and antiparticles
(expressed by subindex $i$) with momentum $\pbf$. We now use the
following \emph{ansatz} for the expression of the
Hamiltonian~(\ref{ham}) as a function of the creation-annihilation operators:
\begin{equation}
\begin{split}
H=& \int \frac{d^3\pbf}{(2\pi)^3} \,\sum_{i}
\left[\ E_i(p)\, b_i^\dagger(\pbf)\,b_i(\pbf)\right. \\ +&
\left. \bar{E}_i(p)\, d_i^\dagger(\pbf)\,d_i(\pbf)\right].
\label{ham2}
\end{split}
\end{equation}
This corresponds to the assumption that the free theory describes
a system of three types of free particles and antiparticles for each
value of the momentum, with energies $E_i(p), \bar{E}_i(p)$, respectively.

Now, computing $[H,\nu_{L\alpha}]$ by two procedures: firstly by using Eq.~(\ref{ham})
for the Hamiltonian and the anticommutators~(\ref{neutrinocomm}), and secondly, by using
the expressions~(\ref{neutrino}) and~(\ref{ham2}), and equalling both results, one
obtains the following simple result for the energies and the coefficients in the
plane wave expansion of the field:
\begin{align}
E_i(p)&=\bar{E}_i(p)=p\,[1+\tilde{c}_i(p)], \label{E} \\
u_{L\alpha}^i(\bm{p})&=v_{L\alpha}^i(\bm{p})=e_\alpha^i(p)\,\chi^i(\bm{p}), \label{uv}
\end{align}
where $\chi^i(\bm{p})$ is the two component spinor solution
of the equation
\begin{equation}
(\bm{\sigma}\cdot \bm{p})\, \chi^i(\bm{p})=-p\, \chi^i(\bm{p})
\end{equation}
with the normalization condition
\begin{equation}
\chi^{i\dagger}(\bm{p})\, \chi^i(\bm{p})=2 E_i(p),
\end{equation}
$\tilde{c}_i(p)$ are the three eigenvalues of
$\tilde{C}_{\alpha\beta}(p)$,
the Fourier transform of $C_{\alpha\beta}(|\bm{x}-\bm{y}|)$ in Eq.~(\ref{neutrinocomm}),
and $e_{\alpha}^i(p)$ are the components of the normalized eigenvectors
of $\tilde{C}_{\alpha\beta}(p)$.

From Eq.~(\ref{E}), we see that the model presented here contains violation
of Lorentz invariance, but preserves $CPT$ symmetry.

\section{New IR physics and neutrino oscillations}

Since in the free theory solution
there are three types of particles and antiparticles with different energies,
and a mixing of creation and annihilation operators of different kinds of
particle-antiparticle in the expression of each field, it is clear that the
nonvanishing anticommutators of different fields will produce neutrino
oscillations, even for massless neutrinos. This observation was
already present in Ref.~\cite{gamboa}.
The probability of conversion of a neutrino of flavor $\alpha$
produced at $t=0$ to a neutrino of flavor $\beta$, detected at time
$t$, can be directly read from the propagator of the neutrino
field (\ref{neutrino}). This probability can be written as
\begin{equation}
{\cal P}\left(\nu_{\alpha}(0)\to \nu_{\beta}(t)\right) \,=\, \left|\sum_i
e^i_{\alpha}(p)^* \, e^i_{\beta}(p) \, e^{-i\,E_i(p)t}\right|^2 .
\end{equation}
This is the standard result for the oscillation between three states
with the unitary mixing matrix elements $U^i_{\alpha}$ replaced by the
coefficients $e^i_{\alpha}(p)$ of the plane wave expansion of the
noncanonical neutrino fields and the energy of a relativistic
particle $\sqrt{p^2 + m_i^2}$ replaced by the energy $E_i(p)$ of the
particle created by the noncanonical neutrino fields. 

Let us now make the assumption that the modification of the anticommutators
is a footprint of new physics at low energies, parametrized
by an infrared scale $\lambda$. Although the introduction of corrections
to a quantum field theory parametrized by a low-energy scale has not been
so well explored in the literature as the corrections produced by
ultraviolet cutoffs, there are several phenomenological and
theoretical reasons that have recently lead to think on the necessity to
incorporate a new IR scale to our
theories~\cite{extradim,gravity,holoQFT,dsrir}.

If the modifications of the anticommutation relations
are parametrized by an infrared scale $\lambda$ then it is reasonable
to assume an expansion in powers of $\lambda^2/p^2$ so that
\begin{equation}
\tilde C_{\alpha\beta}(p)\approx \tilde C_{\alpha\beta}^{(1)}
\,\frac{\lambda^2}{p^2} \text{ for } p^2\gg\lambda^2,
\label{approximation}
\end{equation}
and then
\begin{equation}
\tilde{c}_i(p)\approx \tilde{c}_i^{(1)}\,\frac{\lambda^2}{p^2},
\quad e_\alpha^i(p)\approx e_\alpha^{i(1)},
\label{eigenvalues}
\end{equation}
where $e_{\alpha}^{i(1)}$ ($\tilde{c}_i^{(1)}$) are eigenvectors
(eigenvalues) of $\tilde C_{\alpha\beta}^{(1)}$,
independent of $p$.

But in this approximation, the description of neutrino oscillations produced
by the new physics is completely undistinguishable from the conventional
description based on mass differences ($\Delta m_{ij}^2$) with a
mixing matrix ($U^i_{\alpha}$) between flavor and mass 
eigenstates, just by making the correspondence
\begin{equation}
\Delta m_{ij}^2= 2 \,(\tilde c_i^{(1)}-\tilde c_j^{(1)})\,\lambda^2,
\quad U^i_{\alpha}=e_{\alpha}^{i(1)}.
\end{equation}

One should note that when $\lambda \neq 0$ the different energies for
different states select a basis in the Fock space and the mixing of
Fock space operators in the fermionic fields is unavoidable. It is
only when one considers the energy splitting of the 
different particles that one has a physical consequence of the mixing
of different creation-annihilation Fock space operators in each
fermionic field. On the other hand, in the case $\lambda=0$
(corresponding to unmodified anticommutation relations) there is
an arbitrariness in the construction of the Fock space. One could make
use of this arbitrariness to choose a basis such that each fermionic
field contains only one annihilation and one creation operator (which
is equivalent to saying that the $e_{\alpha}^i(p)$ are indeed the
$\delta_{\alpha i}$) and therefore no oscillation phenomenon is
produced.    

\section{Conclusions}

We have seen in the previous Section that the observations of neutrino
oscillations are compatible with their interpretation as a footprint of
new physics in the infrared. As far as we are aware, this is the first time
that an interpretation of neutrino oscillations coming from new physics,
without the need to introduce neutrino masses, and compatible with
all experimental results, is presented.

This is achieved because of the indistinguishability of the new mechanism
from the conventional one in the range of momenta $p^2 \gg \lambda^2$.
In order to reveal the origin of the oscillations it is necessary to go
beyond the approximation Eq.~(\ref{approximation}), which requires the
exploration of the region of small momenta ($p^2 \approx
\lambda^2$). To get this result it has been crucial to introduce a new
infrared scale through a nonlocal modification of the anticommutation
relations of the neutrino field. Gauge invariance forbids a similar nonlocal
modification for the remaining fields due to the choice of quantum
numbers for the fermion fields in the standard model. In fact the
possibility to have the modified anticommutators (\ref{neutrinocomm}) for the
neutrino fields is related to the absence of the right-handed neutrino
field.

The model of noncanonical fields presented in this work has to be considered
only as an example of the general idea that new infrared physics may be
present in, or be (partially) responsible of, neutrino oscillations, and that
the conventional interpretation may be incomplete. In fact
an extension of relativistic quantum field theory based on the
modification of canonical anticommutation relations
of fields might not be consistent. We have not examined the associated
problems of unitarity or causality beyond
the free theory. But it seems plausible
that the consequences that we have obtained in the neutrino sector will be
valid beyond this specific framework.

In conclusion, in this work we have shown that the experimentally observed
properties of neutrino oscillations do not necessarily imply the
existence of neutrino masses. In fact, future experiments
attempting to determine the neutrino mass, such as KATRIN~\cite{katrin},
may offer a window to the
identification of new physics beyond relativistic quantum field theory
in the IR. At this level it is difficult to predict specific 
observational effects due to the lack of criteria to select a choice 
for $\tilde C_{\alpha\beta}(p)$ in this specific model. A simple
example, however, would be the presence of negative eigenvalues 
of this matrix, which could be reflected in an apparent negative 
mass squared for the neutrino (see Eq.~(\ref{eigenvalues}))
in the fits from the tail of the tritium spectrum.
Effects on cosmology could also be possible, again depending 
on the exact modification of the neutrino dispersion
relation in the infrared. All we can say is that if
the origin of neutrino oscillations is due to new physics in the
infrared then experiments trying to determine the absolute values of
neutrino masses and/or cosmological observations might have a 
reflection of the generalized
energy-momentum relation Eq.~(\ref{E}) for neutrinos.

\acknowledgments
This work has been partially supported by CICYT (grant
FPA2006-02315) and DGIID-DGA (grant2006-E24/2). J.I. acknowledges a FPU
grant from MEC.

\end{document}